\documentclass[apjl,twocolumn]{emulateapj}
\usepackage{epsfig,apjfonts,mathptmx}

\def\gtsima{$\; \buildrel > \over \sim \;$}
\def\ltsima{$\; \buildrel < \over \sim \;$}
\def\prosima{$\; \buildrel \propto \over \sim \;$}
\def\gsim{\lower.5ex\hbox{\gtsima}}
\def\lsim{\lower.5ex\hbox{\ltsima}}
\def\simgt{\lower.5ex\hbox{\gtsima}}
\def\simlt{\lower.5ex\hbox{\ltsima}}
\def\simpr{\lower.5ex\hbox{\prosima}}

\def\h1{$h^{-1}$}
\def\eeq{\end{equation}}
\def\beq{\begin{equation}}
\def\24mu{24\,$\mu{\rm m}$}
\def\70mu{70\,$\mu{\rm m}$}
\def\8mu{8\,$\mu{\rm m}$}

\submitted{ApJ Letters, in press}
\shorttitle{Radio selection of the most distant galaxy clusters}
\shortauthors{Daddi et al.}
\begin{document}

\title{Radio selection of the most distant galaxy clusters}

\author{
E. Daddi\altaffilmark{1},
S. Jin\altaffilmark{1,2},
V. Strazzullo\altaffilmark{3},
M.T. Sargent\altaffilmark{4},
T. Wang\altaffilmark{5,1},
C. Ferrari\altaffilmark{6},
E. Schinnerer\altaffilmark{7},
V. Smol{\v c}i{\'c}\altaffilmark{8},
A. Calabr\'o\altaffilmark{1},
R. Coogan\altaffilmark{1,4},
J. Delhaize\altaffilmark{8},
I. Delvecchio\altaffilmark{8},
D. Elbaz\altaffilmark{1},
R. Gobat\altaffilmark{9},
Q. Gu\altaffilmark{2},
D. Liu\altaffilmark{7},
M. Novak\altaffilmark{8},
F. Valentino\altaffilmark{10}
}

\altaffiltext{1}{CEA, IRFU, DAp, AIM, Universit\'e Paris-Saclay, Universit\'e Paris Diderot,  Sorbonne Paris Cit\'e, CNRS, F-91191 Gif-sur-Yvette, France}
\altaffiltext{2}{School of Astronomy and Space Science, Nanjing University, Nanjing 210093, China}
\altaffiltext{3}{Department of Physics, Ludwig-Maximilians-Universitat, Scheinerstr. 1, 81679, Munchen, Germany}
\altaffiltext{4}{Astronomy Centre, Department of Physics and Astronomy, University of Sussex, Brighton, BN1 9QH, UK}
\altaffiltext{5}{Institute of Astronomy, the University of Tokyo, and National Observatory Of Japan, Osawa, Mitaka, Tokyo 181-0015, Japan}
\altaffiltext{6}{Universit\'e Cote d'Azur, Observatoire de la Cote d'Azur, CNRS, Laboratoire Lagrange, Bd de l'Observatoire, CS 34229, 06304 Nice cedex 4, France}
\altaffiltext{7}{Max Planck Institute for Astronomy, Konigstuhl 17, D-69117 Heidelberg, Germany}
\altaffiltext{8}{Department of Physics, Faculty of Science, University of Zagreb, Bijeni{\v c}ka cesta 32, 10000, Zagreb, Croatia}
\altaffiltext{9}{School of Physics, Korea Institute for Advanced Study, Hoegiro 85, Dongdaemun-gu, 02455, Seoul, Republic of Korea}
\altaffiltext{10}{Dark Cosmology Centre, Niels Bohr Institute, University of Copenhagen, Juliane Maries Vej 30, DK-2100 Copenhagen, Denmark}

\begin{abstract}
We show that the most distant X-ray detected cluster known to date, Cl~J1001 at $z_{\rm spec}=2.506$, hosts a strong overdensity of radio sources. Six of them are individually detected (within 10$''$) in deep 0.75$''$ resolution VLA 3GHz imaging,  with $S_{\rm 3GHz}>8\mu$Jy. Of the six, AGN likely affects the radio emission in two galaxies while star formation is the dominant source powering the remaining four.
We searched for cluster candidates over the full COSMOS 2-square degree field using radio-detected 3GHz sources and looking for peaks in  $\Sigma_5$  density maps. Cl~J1001 is the strongest overdensity by far with $>10\sigma$,  with a simple $z_{\rm phot}>1.5$ preselection. A cruder photometric rejection of $z<1$ radio foregrounds leaves Cl~J1001 as the second strongest overdensity, while even using all radio sources Cl~J1001 remains among the four strongest projected overdensities. We conclude that there are great prospects for future, deep and wide-area radio surveys to discover large samples of the first generation of forming galaxy clusters. In these remarkable structures widespread star formation and AGN activity of massive galaxy cluster members, residing {\em within} the inner cluster core, will ultimately lead to radio continuum as one of the most effective means for their identification, with detection rates expected in the ballpark of 0.1--1 per square degree at $z\simgt2.5$. Samples of hundreds such high-redshift clusters could potentially  constrain cosmological parameters and test cluster and galaxy formation models.
\end{abstract}

\keywords{galaxies: evolution --- galaxies: formation ---   galaxies: clusters: general ---   radio continuum: galaxies --- galaxies: high-redshift }

\section{Introduction}

The identification of the most distant, $z\simgt2$ galaxy clusters --  i.e. structures consistent with a single massive dark matter halo as opposed to Mpc-scale loose overdensities like proto-clusters (Diener et al. 2015)    -- is a very active topic of current research (see Overzier et al. 2016 for a recent review). 
They represent the earliest generation of massive collapsed structures, progenitors to Coma-like clusters, and their abundance can constrain cosmological parameters. Also, they are unambiguous  formation sites of massive ellipticals, and hence hold promise to shed light on the elusive processes that lead to the formation of passive, early-type galaxies. It is currently unclear if the morphological transformation of galaxies into spheroidal systems happens before or after entering the densest early cluster cores. The same holds for the quenching of star formation and passivisation. 
High redshift clusters are also interesting laboratories to study (Valentino et al. 2016) the interaction between star-formation/AGN activity and the hot intra-cluster medium (ICM), 
energy injection into the ICM and its thermodynamical evolution. At redshifts $z\sim$1.5--3 theory predicts that massive $\sim10^{13-14}M_\odot$ dark matter halos should undergo a transition from being fed by cold streams to being shielded  by a hot atmosphere which prevents refueling with fresh gas  (Dekel et al. 2009). This process is not well constrained by theory. Observations are needed to trace the redshift and duration of this transition. Evidence of persistent activity in massive structures at $z\sim2$--2.5 suggests that this might occur later and/or at higher masses than currently expected (Valentino et al. 2015; 2016; Wang et al. 2016; Overzier et al. 2016). Statistics from a larger number of clusters are required for definitive conclusions. 
Surveys aiming at detecting hot ICM, via X-rays or Sunyaev-Zeldovich (SZ) emission, generally lack sensitivity for identifying clusters beyond $z\sim2$. Clusters found approaching such a limit are so massive ($>>10^{14}M_\odot$; Newman et al. 2014; Stanford et al. 2012) and evolved (i.e., dead) to be far less interesting for the science discussed above. More typical $z\simgt2$ systems have been discovered looking for concentrations of massive galaxies  (e.g., Gobat et al. 2011; 2013;  Yuan et al. 2014; Strazzullo et al. 2015; Wang et al. 2016), and X-ray emission searched for at their known positions, a posteriori. 

\begin{table*}
{
\caption{Radio detections in the W16 cluster}
\label{tab:1}
\centering
\begin{tabular}{ccccccccrrlr}
\hline\hline
     ID-Jin & ID-COSMOS2015   &      RA      &    DEC     &   distance      &  z$_{\rm phot}$    &  z$_{\rm spec}$ &  log M$^*$  &  S$_{\rm 3GHz}$ &   S$_{\rm 850\mu m}$ &       Ks & Origin of radio \\
               & or Muzzin		  &       deg         &     deg        &     $''$          &           &     &  $M_\odot$ & $\mu$Jy & mJy  &  AB mag & \\
                  \hline
681633 &  130651    & 150.2389798 & 2.3339305    &    8.5    &    2.69    & $2.50\pm0.02$ &  11.18    &     27.6  $\pm$       2.7      &   $< 1.2$ (3$\sigma$)  &      22.43 & AGN \\
683281 & 130891    & 150.2398843 & 2.3364591    &    4.9   &     2.74    &   2.513 & 11.33    &     20.6    $\pm$     2.5      &  3.77 $\pm$ 0.32   &     22.56 & SB \\
683795 & 130933     & 150.2387007 &  2.3368268   &      2.2  &      2.29    &   2.500 & 11.02    &     24.1    $\pm$     2.7       &   1.66 $\pm$ 0.21  &    22.31 & AGN \\
684410 & 130901     & 150.2392700 & 2.3363813    &    2.7  &      2.36    &   2.508 & 11.23    &    8.8   $\pm$    2.9    &   2.23 $\pm$ 0.41      &   21.7 & MS \\
684496 & 130949    & 150.2370141 & 2.3357152    &    5.8  &      2.26    &   2.503 & 11.32    &    10.8   $\pm$    3.2      &   1.69 $\pm$ 0.25   &   20.93 & MS \\
10131077 & 131077    & 150.2373500 & 2.3381379   &     8.0    &    2.82  &  2.494 &    10.92   &      39.5    $\pm$     3.1     &   5.26 $\pm$ 0.26 &          24 & SB \\
\hline
\hline
\end{tabular}\\
}
{Notes: The photometric redshifts, coordinates and Ks magnitudes are from the COSMOS2015 catalog (Laigle et al. 2016), except   ID-10131077  (Muzzin et al. 2013), with Ks measured at the radio position. Spectroscopic redshifts are from W16.
Radio measurements and relative IDs are from S. Jin et al. (2017, in preparation) catalog. Distances are relative to the cluster centre which is located at
[150.2385331,                 2.3362418]. For the 'Origin of radio' column see the discussion in the text.}
\end{table*}

\begin{figure*}[ht]
\centering
\includegraphics[width=18cm]{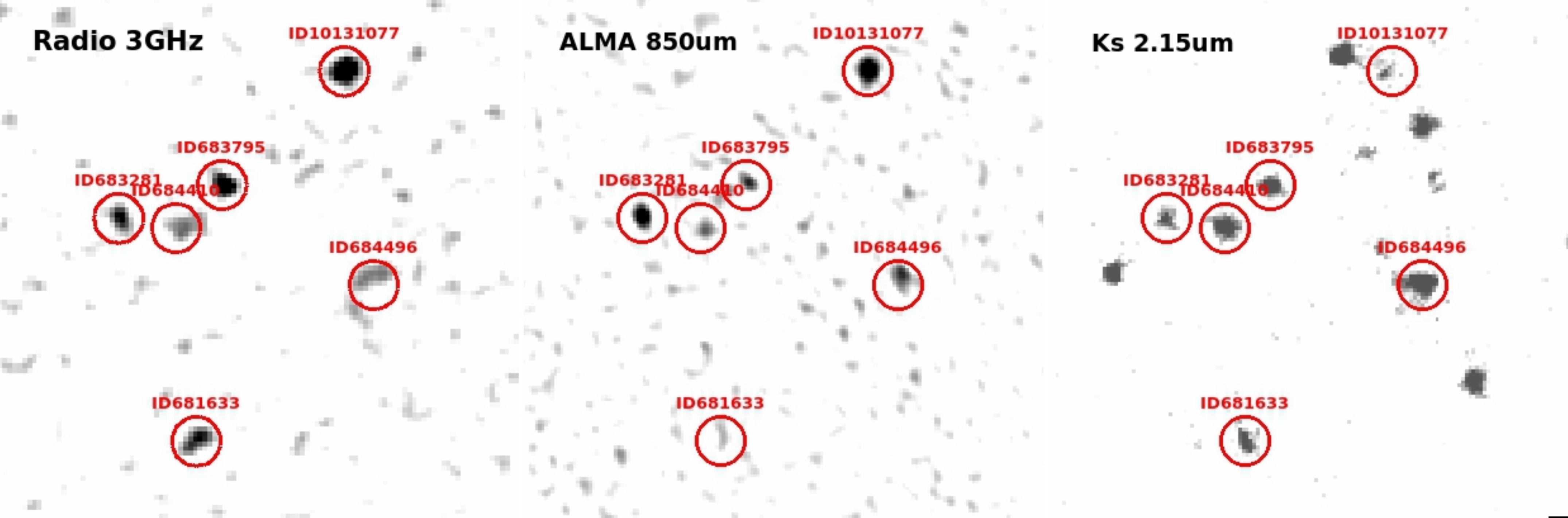}
\caption{Maps of the W16 cluster region: VLA 3GHz (left), ALMA 850$\mu$m (center), Ks-band (right), all $\sim0.7''$ resolution. Panels are 20$''$.
Small circles (1'' radius)  help localizing radio detected cluster members (all spectroscopically confirmed, see Table~1).}
\label{fig:1}
\end{figure*}

Quite strikingly, the most distant X-ray detected cluster known to date at $z=2.506$ (Wang et al 2016;  W16 hereafter) displays a startling amount of star formation in its core ($L_{\rm IR\ 8-1000\mu m}\sim2\times10^{13}L_\odot$). Searching for concentrations of galaxy activity from star formation or AGN might thus be a competitive means for finding the most distant structures. In this letter we explore this idea analysing the W16 cluster radio continuum properties, having in mind that forthcoming deep radio surveys will cover large sky areas to remarkable depths. We use standard cosmology $(70, 0.3, 0.7)$ and a Chabrier IMF.

\section{Radio continuum emission from galaxies in a $z=2.5$ cluster}
\label{sample}

Deep VLA observations of the COSMOS field at 3GHz have been obtained, fully reduced, analysed, and publicly released by Smol{\v c}i{\'c} et al. (2017). 
We have used the PSF fitting technique and cataloging method of Liu et al. (2017) to obtain radio flux density measurements at 3GHz for all Ks-selected galaxies 
in the COSMOS2015 catalog of Laigle et al. (2016), supplemented with sources from Muzzin et al. (2013). Full details of this radio catalog together with the multi-wavelength properties in {\em Herschel}, Spitzer, SCUBA2 and other submm probes of COSMOS galaxies will be presented in a forthcoming publication (S. Jin et al. 2017, in preparation). We concentrate here on the radio properties of galaxies in the surroundings of the W16 cluster, as compared to those in the full COSMOS 2-square degree field. Thanks to our PSF fitting technique we can push to deeper radio flux levels compared to blindly extracted $S/N>5$ catalogs in Smol{\v c}i{\'c} et al. (2017), with high fidelity and completeness and with very low expected spurious detection rate. Our simulations  return  typical rms sensitivities of 2.5--2.7$\mu$Jy with well behaved gaussian-like uncertainties, close to the expected 2.3$\mu$Jy noise, allowing for reliable S/N$>3$ detections at 3GHz down to about 8$\mu$Jy.
Radio sources without a Ks counterpart would be lost by our technique, and are added back to the sample from the Smol{\v c}i{\'c} et al. catalog 
(sources with multiple cataloged components, e.g. radio lobes, had been combined into one catalog entry).

\begin{table*}
{
\caption{Evaluation of overdensity of the W16 cluster from radio plus ancillary selections}
\label{tab:2}
\centering
\begin{tabular}{cccccll}
\hline\hline
Selection   & Full sample & \# in $R=10''$(1)   & P($\geq 6$)(2) &  Det. rate(3)  & $\Sigma_5$ peak rank & Comments(4) \\
                  \hline
S/N$_{\rm 3GHz}>3$ & 17803 & 0.25 & 3.0$\times10^{-7}$ & 0.02 & 4 & 7.6$\sigma$, after 3 with 7.8--8.1$\sigma$ \\
S/N$_{\rm 3GHz}>3$ \& $Ks>20.5$ & 8016 & 0.11 & 2.8$\times10^{-9}$ & $2.0\times10^{-4}$  & 2 & 8.7$\sigma$, after a $9.2\sigma$ \\
S/N$_{\rm 3GHz}>3$ \& $z_{\rm phot}>1.5$ & 4952 & 0.07 & 1.6$\times10^{-10}$ & $10^{-5}$  & 1 & $10.0\sigma$, the second is at $7.6\sigma$ \\
S/N$_{\rm 3GHz}>3$ \& $2<z_{\rm phot}<3$ & 1994 & 0.03 & 1.4$\times10^{-12}$ &  $10^{-7}$ & 1 & $12.3\sigma$, the second is at $6.5\sigma$ \\
\hline
\hline
\end{tabular}\\
}
{Notes: (1) assumes 1.7 square degrees area of the Laigle Ultra-VISTA catalog. (2) Poisson probability. (3) Number of expected chance associations of $\geq6$ objects within 10$''$ in the full COSMOS field, neglecting clustering effects, after considering there are 70000 independent realizations of 10$''$ radius fields, in COSMOS. (4) The significance (in $\sigma$) is computed respect to the peak value of the distribution of $log(\Sigma_5)$ readings over the whole field. The $\sigma$ is defined as the squared root of the variance of the best fitting $\Gamma$ distribution. The variance is in all cases $\sim0.04$ in the log, while the field average moves from  $10^4$~deg$^{-2}$ for the full radio sample to $10^3$~deg$^{-2}$ for the  $2<z_{\rm phot}<3$ case (see also Fig.~3, the average separation changing by a factor of 3 as a result of the changing number density).
}
\end{table*}

Fitting a total of 589713 priors we obtain 17803 radio continuum detections with S/N$>3$ in the COSMOS field,  extracted over an area of 1.7~deg$^2$ from our Ks-band priors (doubling the number of detections at fixed area, compared to the blind catalog). We find that six galaxies are radio detected within a 10$''$ radius from the center\footnote{Average position of the 6 radio sources.} of the W16 cluster, see Fig.~1 and Table.~1. These are all spectroscopically confirmed cluster members and very red galaxies. The faintest, two MS galaxies, are not present in the Smol{\v c}i{\'c} catalog. Four of the radio detections exhibit radio flux densities consistent with the FIR-radio correlation for star-forming galaxies at their redshift (Yun et al. 2001;  Delhaize et al. 2017). Their radio emission is naturally explained by star-formation alone.
Two of these are relatively low mass while hosting high SFRs, i.e. they are starburst-like galaxies, probably merger-driven. The other two are consistent with the main sequence at $z=2.5$ (see also W16). The remaining two detections have a radio flux excess over the SFRs inferred from ALMA (Table~1; W16) by factors of $\simgt$3--5   and do likely contain  radio AGNs, with moderate intrinsic luminosities $\simlt10^{25}$~W~Hz$^{-1}$  (at rest 1.4~GHz, using also  1.4~GHz data;  Schinnerer et al. 2010). These two radio AGN do not exibit any signature of mid-IR or X-ray AGN activity (e.g. Delvecchio et al. 2017). The median  ALMA 850$\mu$m to radio 3GHz  flux ratio is $\sim150$ (Table~1).

The total radio flux of the 6 cluster detections is $130\mu$Jy with PSF fitting, or max. 170$\mu$Jy when accounting for the possibly extended nature of some sources. Depending on the exact value, which radio-IR correlation we adopt, and factoring in possible AGN components, this corresponds to 20--90\% of the expectation from their {\em Herschel}+ALMA luminosities (W16). We radio-detect all sources producing the bulk of the SFR activity, with tentative evidence of a lower radio/L$_{\rm IR}$ ratio compared to the field.

Fig.~\ref{fig:2} shows distributions of photometric redshift, Ks-band magnitudes and radio fluxes. Based on Kolmogorov-Smirnov tests, the $z=2.5$ cluster galaxies are significantly fainter in Ks and at higher redshift than most radio detections above 8$\mu$Jy at 3GHz. Instead, their radio flux density distribution, spanning 8--40$\mu$Jy (see Tab.~1), is still consistent with a random sampling of the parent catalog.

\section{Radio selection of the $z=2.5$ W16 cluster}

The 6 detections single out the W16 cluster as a very special environment in the radio, enough to allow its pre-selection as a cluster candidate even based on radio information alone. Such selection can reach even stronger confidence when some low-level ancillary information is used in addition to radio, such as could be expected to be available in support of future generation, wide area radio surveys. Following Fig.~\ref{fig:2}, we considered 
the case of near-IR Ks-band imaging being available, or some moderate-quality photometric redshifts  distinguishing $z>1.5$ sources,
like the case for example with the BzK color preselection (Daddi et al. 2004) or red color selections (Franx et al. 2003; see also W16) or IRAC/WISE red colors (e.g., Papovich et al. 2010). Finally, we also considered the direct use of COSMOS-quality photometric redshifts with an accuracy of about 7\% in $(1+z)$ at $z\sim2$--3 as available for W16 cluster galaxies.

As a first step, we compute the Poisson probabilities for chance associations of 6 galaxies within $10''$ radius. This approach is approximative, given that the size and radius of the search area were chosen ad hoc, although we note that the 10$''$ radius   corresponds to the core radius of a massive $\sim10^{14}M_\odot$ cluster at $z>2$ (W16). More importantly, this approach neglects clustering, i.e. presence of galaxies in smaller structures, filaments and even projected random association of physical galaxy pairs, that can conspire to increase the spurious (non single cluster) occurrence of a large counting event. This can nevertheless provide a useful starting guide.
The Poisson probability of a W16-like association is  fairly low even when  just considering  radio sources: 0.02  events are expected in COSMOS. This decreases to $2.0\times10^{-4}$ considering only Ks-faint radio sources, and to $10^{-5}$ and $10^{-7}$ for sources with $z_{\rm phot}>1.5$  and $2<z_{\rm phot}<3$, respectively.

\begin{figure*}[ht]
\centering
\includegraphics[width=18cm]{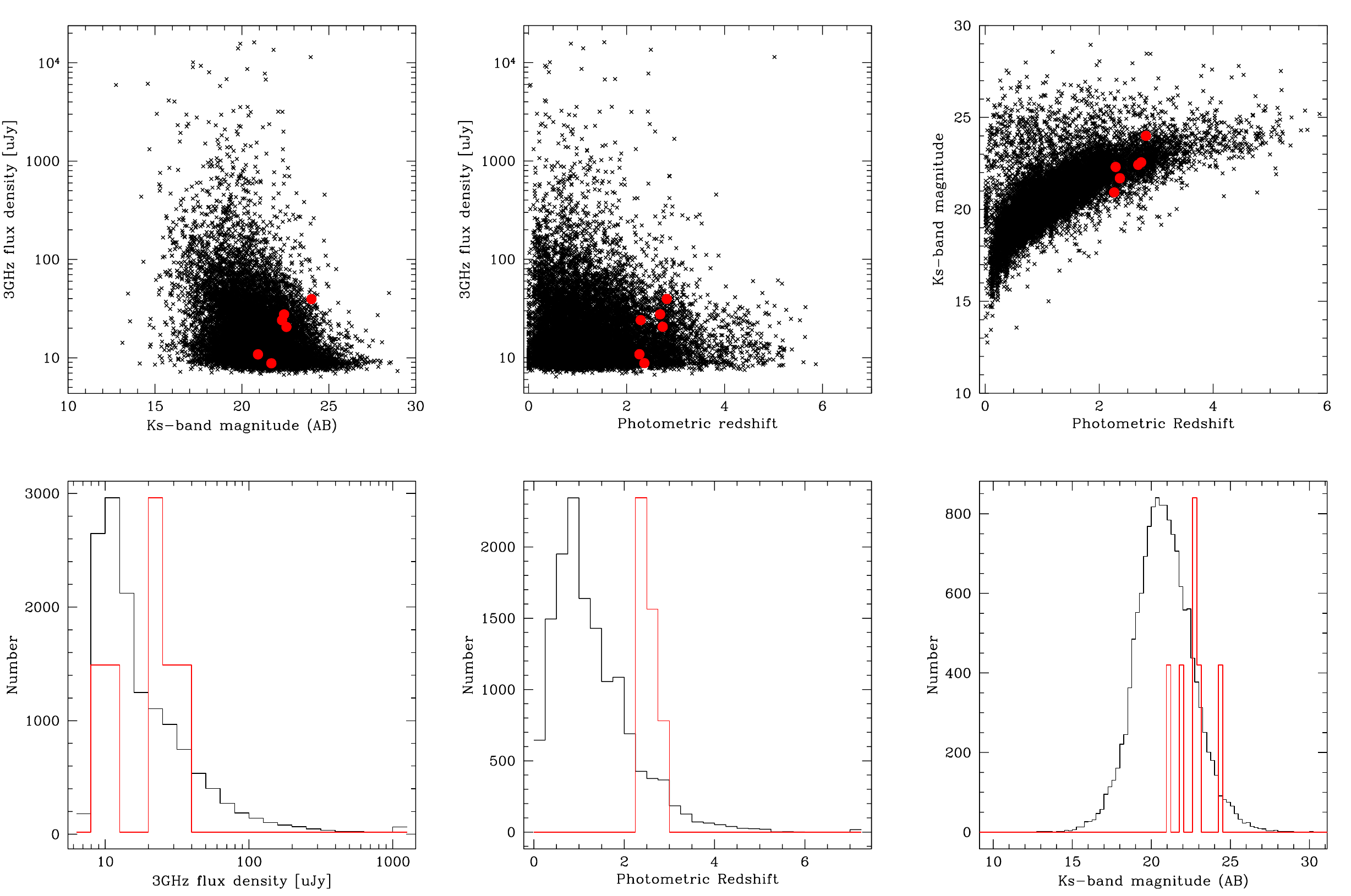}
\caption{Observed properties, including radio fluxes, Ks-band magnitudes and photometric redshifts, of the radio detections in the W16 cluster (red) compared to the parent sample of all 3GHz detections in COSMOS. Red histograms values are arbitrarily scaled to fit the plot.
}
\label{fig:2}
\end{figure*}

A more objective approach is to use  $\Sigma_5$ peak statistics\footnote{$\Sigma_5=5/(\pi R_5^2)$, where $R_5$ is the distance to the 5th closest neighbor. Also this approach requires choosing a number (5). An alternative approach would be convolving the catalog with a 2D profile as expected for a cluster, see Fig.9 in W16. We find this  leads to consistent results.} to search for overdensities in the full COSMOS field, following the method from Strazzullo et al. (2015) which allows for a derivation of the dispersion and thus for an evaluation of their significance.  
Fig.~3 shows the resulting $\Sigma_5$ maps and distribution of $\Sigma_5$ readings over the full COSMOS areas for the various selections described in Tab.2. When allowing for a photometric redshift pre-selection of high redshift radio sources, the W16 cluster is recovered as the first ranked overdensity. Pre-selecting radio sources with $2<z<3$  the W16 cluster is a 12.3$\sigma$ overdensity, long ahead of the second ranking object, seen with 6.5$\sigma$ at position [149.86924, 2.3417165]. With a cruder $z_{\rm phot}>1.5$  constraint, as obtainable from color selections, the W16 cluster is detected at $>10\sigma$. The second ranking cluster is again detected with far less significance at $7.6\sigma$ near [150.31505,  2.7126151]. This latter structure seems loose and possibly the result of line-of-sight alignments. Using all Ks-faint radio sources returns the W16 cluster as the second ranking overdensity with $8.7\sigma$, after a $9.2\sigma$ at [150.14933,  2.5922460] (however the latter is likely a chance alignment including low-z sources). 
When using all radio detected sources with $S/N>3$ the W16 is found as the 4th ranked ($7.7\sigma$), after 3 more 7.8--8.1$\sigma$ peaks
[150.14933, 2.5933659]-[150.00879,2.2741797]-[149.83185, 2.5704069]. All of the latter  appear to be chance alignments, judging from their $z_{\rm phots}$, except perhaps a possibly genuine but loose concentration at $z\sim0.9$ in the latter case. A detailed discussion of the nature of these possibly significant $\Sigma_5$ peak is beyond the scope of this letter, where we concentrate on the well studied case of the spectroscopically confirmed, $z=2.506$ cluster W16.

\begin{figure*}[ht]
\centering
\includegraphics[width=18cm]{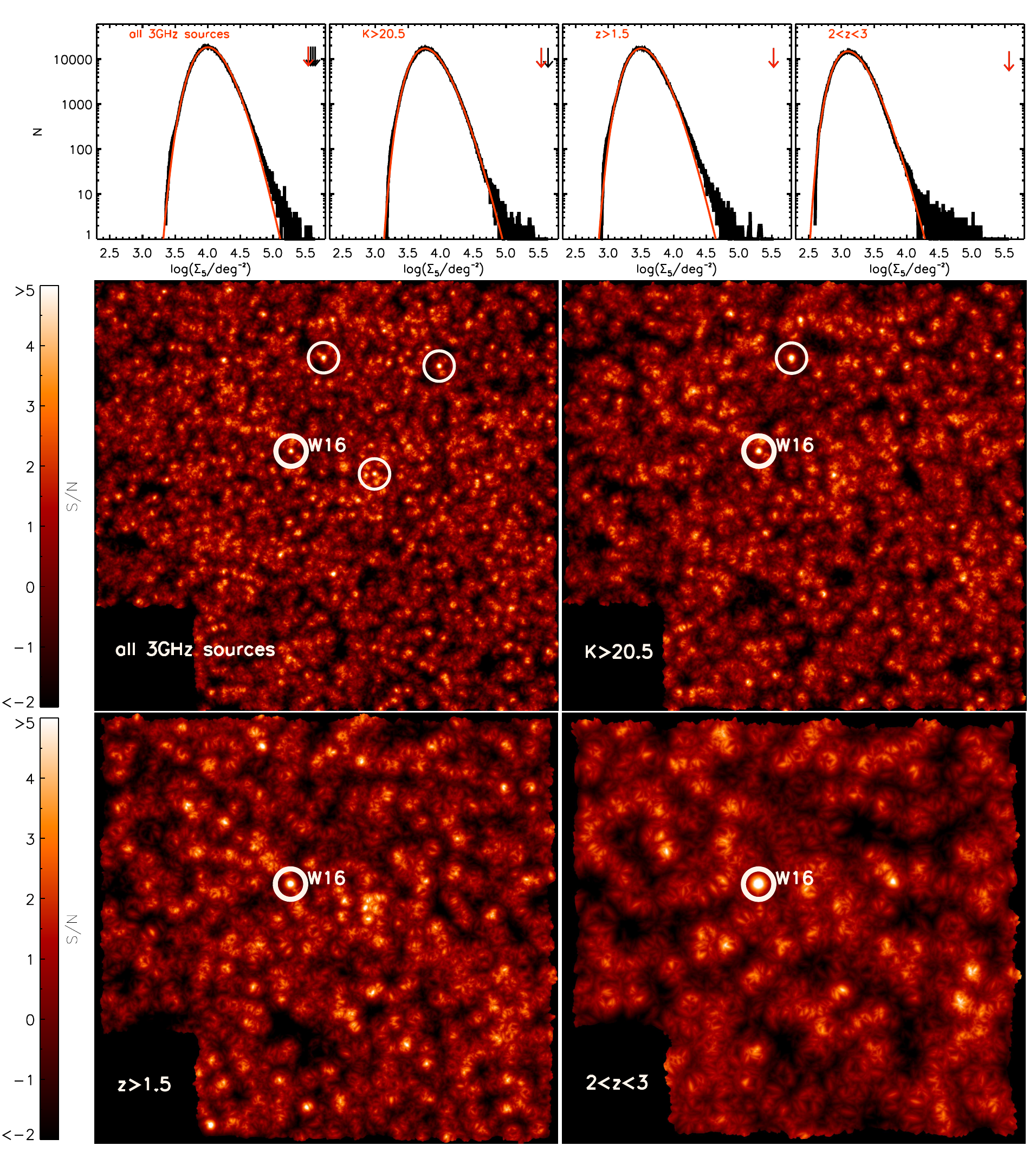}
\caption{$\Sigma_5$ maps (bottom four panels) and distributions of readings (top four panels),  for the sample selections discussed in the text. The W16 cluster is shown with a thicker white circle uniquely identified in the 2 bottom maps. The bottom-left corner of each map is not analysed due to poor coverage in the Ks-band. The histograms are fitted with $\Gamma$ distributions. Arrows show detected overdensities down to the W16 cluster (red). Fig.1 of W16 provides a useful comparison of a similar diagnostic on Ks-selected galaxies. 
}
\label{fig:3}
\end{figure*}

We caution that with 6 sources only and with two close to the limit  of our current radio photometry, the W16 cluster provides a highly significant but perhaps fragile detection. One could  imagine that out of many realisations of a similar dark matter halo structure at $z=2.5$, at times one or two galaxies could remain below current detection limits (for effects of noise, lower SFR, lower AGN activity), preventing detection in the radio in presence of low number statistics. 
Those four of the six W16 radio detections which appear in the Smol{\v c}i{\'c} et al. 3 GHz catalog (i.e. have $S_{\rm 3GHz}>20\mu$Jy) still define an $\sim11\sigma$ overdensity (when adopting, ad-hoc, the $\Sigma_4$ statistics).

\section{Discussion}

We have demonstrated above that using only radio detected sources, ideally with alternative information in the form of photometric redshifts or even single band photometry, allows us to pinpoint the W16 cluster. This is quite promising news for future large area radio surveys, which will reach very faint flux densities (Padovani 2016). For example,  even in phase1 SKA will likely carry out a full sky survey to few $\mu$Jy sensitivity and with $\sim2''$ resolution at 1--2~GHz, while covering $\sim1000$ deg$^2$ to $\sim1\mu$Jy depth and sub-arcsec resolution (Prandoni \& Seymour 2015),  surpassing the depth and sensitivity of our 3GHz VLA data, allowing in principle to find W16-like clusters over the whole sky, and perhaps much further away, and in great numbers. Current plans envisage excellent prospects also for ngVLA, when observing at $\sim$2--3~GHz (Selina \& Murphy 2017). Our work identifies important synergies between future radio, optical, NIR and X-ray surveys. The community preparing future multi-band facilities is perfectly aware of the necessity of this joint effort, which therefore starts to be discussed and organised (see e.g. Table 5 in Prandoni \& Seymour 2015; see also Bacon et al. 2015 [LSST vs SKA], Kitching et al. 2015, Ciliegi \& Bardelli 2015 [Euclid vs. SKA]). For example, Euclid full-sky photometry (compare to Fig.~2) should be already sufficient to reproduce the different galaxy selections that we explored in this work, thus likely greatly enhancing the contrast towards genuine high redshift cluster structures found in the radio.  On the other hand, radio surveys over large areas but with shallower depth and/or lower angular resolution than the 3GHz COSMOS data (we recall we have 0.75$''$ resolution and 2.3$\mu$Jy rms at 3GHz, equivalent to 4.3$\mu$Jy at 1.4GHz for $\alpha=0.8$) like the EMU survey with ASKAP at 1.4 GHz, LOFAR low frequency surveys, and even MIGHTEE  from MeerKAT -- see details in Norris et al. 2013 for all of these -- would be substantially less efficient in finding $z>2.5$--3 clusters. 

ALMA interferometric imaging (Fig.1 and Tab.1) will provide efficient screening of radio-selected high-z cluster candidates,
 confirming the high galaxies SFRs and their distances (the radio to submm flux ratio is a rough redshift indicator -- Carilli \& Yun 1999). The W16 cluster is  the strongest SPIRE  source in the {\em Herschel} imaging of CANDELS-COSMOS. While from {\em Herschel} alone it is impossible to know if that corresponds to a single bright starburst or multiple sources,
this is within the reach of ALMA that can deliver sub-arcsec resolution in the submm.
With less than one minute per field at 345GHz  ALMA  can nowadays reach deeper than the data presented in Tab.1 and Fig.1, making  ALMA pre-screening of high-z cluster candidates from radio surveys viable even for large samples of cluster candidates. The 20$''$ ALMA field of view at 345GHz  is  well matched to the size of the overdensities, and W16 already demonstrated that millimeter spectroscopy with NOEMA/ALMA/VLA is a competitive means to spectroscopically confirm the clusters in a single shot.
On the other hand, ALMA cannot perform blind cluster searches on large areas. Observations at lower frequencies in the radio have substantial advantages with much wider fields of view. 

Overall, it appears that the reason why there is such a strong contrast in radio for the W16 cluster is twofold: first, the overall strong rise of specific SFR to high-z  (e.g., Schreiber et al 2015; Faisst et al. 2016) with its accompanying AGN activity (e.g., Mullaney et al. 2012) imply fairly high radio fluxes for massive star forming galaxies. But even more crucially, the vast majority ($>80$\%) of massive galaxies in the W16 cluster core are strongly star-forming ($SFR\simgt100M_\odot$~yr$^{-1}$) rather than passive (the opposite is seen even in the $z\sim2$ cluster Cl~1449; Gobat et al. 2013; Strazzullo et al. 2013; see also Newman et al 2014). 
This appears to be a genuine transition to what could be objectively defined as a {\em star forming cluster}. Such star forming clusters would be deliverable in large numbers by  radio 
pre-selection in future radio surveys, as advocated here. As discussed in the Introduction, these would be some of the most interesting places to study early galaxy formation and evolution 
in cluster structures, with the cluster red sequence still in the process of birth at $z\sim2$ (Strazzullo et al. 2016), and the interplay between ICM and galaxy formation (Valentino et al. 2016). 
Quite interestingly, visual inspection suggests elongated radio morphology (misaligned with the optical) in 3 of the 6 detections in the W16 cluster (IDs 681633 [AGN], 684496 [SF], 684410 [SF], see Fig.~1), hinting to the possibility that small-scale radio jets might be present in half of the sources. 
Radio might thus provide insights into cluster physics and interaction between galaxies and ICM. SKA1 band2 should provide 0.4$''$ resolution and deeper data over an area  $\times1000$ times larger than COSMOS, and will clarify this issue.

The W16 cluster recovery  performance obtainable from radio samples is quite comparable to that achievable using $J-Ks$ color on Ks-selected samples, that returned $11.6\sigma$  (W16) respect to 12.3$\sigma$ here,  both 1$^{\rm st}$ rank. The two techniques are complementary and could beneficially be used together: the  advantage of the radio is that it is unsensitive to dust obscuration and directly selects on the galaxy star-formation activity (and AGN emission adds on top of that) thus pinpointing forming/active clusters, while a pure selection on optical colors could potentially reveal old and dead clusters as well.  This approach is also quite different from the use of very luminous radio-galaxies  as massive  beacons to locate distant clusters (e.g. Miley et al. 2006; Miley \& deBreuck 2008; Galametz et al. 2012; Cooke et al. 2015), but future large area radio surveys might combine the two efforts, allowing one to select at the same time the rare ultra-bright radio galaxies as well as the faint cluster galaxies, thus clarifying the cosmic relevance of (proto-)clusters around radio-galaxies. While subject to similar limitations as  {\em Herschel} searches, Planck Collaboration XXXVII  and  XXXIX (2015; 2016) selected large samples of high redshift proto-cluster candidates. However, these candidates  mostly  comprise chance alignments of smaller structures at different redshifts (Negrello et al. 2017).

Based on the recovery of a single W16-like cluster in the COSMOS field, we expect a return of 0.1--1~deg$^{-2}$ W16-like clusters at the 68\% confidence level (Poisson counts).  Even limiting to the SKA1  survey over 1000~deg$^2$ could thus produce between 100-1000 $z\simgt2.5$ clusters. 
The number density of similarly massive halos at $z>3.5$ would be about $\times10$ smaller, still allowing sizeable samples of up to 10--100.  
At $z=3.5$ we would expect $\times2$ smaller radio fluxes compared to $z=2.5$ at fixed observing frequency and luminosity. This could be mitigated by the expected increase of the specific SFR with redshift, and the SKA1 will reach deeper than our current data.

As discussed in W16, the identification of this structure in a field like COSMOS is somewhat in tension with $\Lambda$CDM, suggesting that actual returns will be towards the lower range, or the mass of the W16 system overestimated. However, expected numbers are large enough that future samples of high redshift clusters might allow interesting tests of cosmology, in addition to cluster and galaxy formation physics, given that Athena will have the sensitivity to constrain their dark matter halo masses from the X-rays (e.g., Padovani et al 2017).  

\begin{acknowledgements}
SJ acknowledges Chinese Scientific Council funding. MTS was supported by a Royal Society Leverhulme Trust Senior Research Fellowship (LT150041). We acknowledge  European Union's Horizon 2020 support (ERC 694343: ES, DL;  ERC 337595:  VS, JD, ID, MN). QG acknowledges the National Key Research and Development Program of China (No. 2017YFA0402703).
\end{acknowledgements}

\end{document}